\documentclass[twocolumn,prl,floatfix]{revtex4}

\usepackage{color}
\usepackage{epsfig}
\begin{document}
\title{Elastic domains in antiferromagnets}
\author{Efim A. Brener$^*$ and V.I. Marchenko$^{**}$\\
$^*$Institut f\"ur Festk\"orperforschung, Forschungszentrum
J\"ulich, D-52425 J\"ulich, Germany\\
$^{**}$P.L. Kapitza Institute for Physical Problems, RAS
119334, Kosygina 2, Moscow, Russia}
\date{\today}
\begin{abstract} We consider 
 periodic domain structures which appear due to the magnetoelastic interaction 
if the antiferromagnetic crystal is attached to an elastic substrate.
 The peculiar behavior of such structures in an external magnetic field is discussed.
In particular, we find the magnetic field dependence  of the  equilibrium period 
and the concentrations of different domains.
\end{abstract}
\maketitle

Spontaneous breaking of any discrete rotational symmetry in
monocrystals leads to the appearance of  degenerate phases
characterized by different orientations of the order parameter. The
phase boundaries have a positive (free) energy, otherwise the system
would be thermodynamically unstable. Thus, in equilibrium one should
observe a single phase. This is, however, not the case for the phases with
spontaneous magnetization (or electrical polarization). 
 Periodic domain structures are formed in such materials in order to
reduce the large, proportional to the  volume, energy of the magnetic (or
electric) field.

Using the electron spin resonance technique,
Janossy {\it et al.} \cite{J99,J03}  demonstrated the existence
of the equilibrium domain structure in antiferromagnetic $YBa_2Cu_3O_{6+x}$ and 
in $Y_{1-x}Ca_xBa_2Cu_3O_6$.   
The domain structure was easily
 modified by the external magnetic field, 
and restored after switching  the field off.
The resonance method does not provide  information about characteristic sizes and 
the arrangement of the domains. Recently Vinnikov, {\it et al.} \cite{V03} directly 
observed periodic domain structures in antiferromagnetic $TbNi_2B_2C$ 
by the ``finest magnetic-particle decoration technique'' \cite{V93}.   Neither
spontaneous magnetization, nor electrical polarization is seen in this material. 
Nevertheless, regular and reversible 
(by the change of the temperature and magnetic field)
domain structures are realized with a periodicity of the order of a few microns
in the samples of  plate shape with a thickness 0.5 mm.

In this Letter we discuss a possible scenario for the appearance
of such periodic domain structures
and the  behavior of the structures in an external magnetic field.
If a monocrystal is attached to some elastic substrate, 
the domain structure should inevitably  arise
in any  orientational phase transition in order to minimize the
strain energy. Indeed,
the stress free (but not strain free)  monodomain state is realized
if we discuss  free surface boundary conditions. If the monodomain crystal
is attached to some elastic substrate,  stresses arise, and we
loose  a large, proportional to the crystal volume, energy. In this case, the  
appearance of the domain structure, with a period much smaller
than the crystal size, will  drastically diminish the energy of elastic deformations.

\begin{figure}
\begin{center}
\epsfig{file=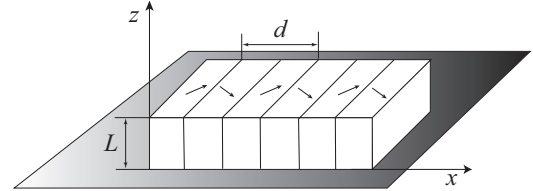, width=8cm}
\caption{Schematic presentation of the domain structure in the antiferromagnetic
crystal attached to an elastic substrate.}
\end{center}
\end{figure}
All materials studied in the experiments of Ref. \cite{J99,J03,V03} 
are collinear easy plane  antiferromagnets with  tetragonal crystal symmetry. 
For definiteness' sake  we discuss this case. 
For the description of the antiferromagnetic 
states we use the exchange approximation \cite{AM}, treating the effects 
of the magnetic
anisotropy, magnetoelastic effects and the external magnetic field as perturbations. 
By a proper choice of the orientation  of the coordinate system (say, 
the $x$ axis can be oriented along the [110] or [100]
direction) we can write the anisotropy energy in the easy 
$xy$-plane as
\begin{equation}\label{an}
F_{an}=-\beta l_x^2l_y^2,
\end{equation}
with a positive constant ${\beta}$; here $l_x,$
and $l_y$ are the components of the antiferromagnetic unit vector ${\bf l}$. 
The four states ${l^2_x=l^2_y=1/2},$ differ by the signs of the
 $x,y$-components $(\nearrow,\, \searrow,\, \swarrow,\, \nwarrow),$ and
 correspond to the minimum of the energy, Eq. (\ref{an}).

The elastic strain arises in the crystal due to the magnetoelastic
interaction
\begin{equation}\label{me}
F_{me}=\gamma_1l_xl_yu_{xy}+\gamma_2(l^2_x-l^2_y)(u_{xx}-u_{yy}),
\end{equation}
where $u_{ik}$ are the components of the strain tensor. For simplicity, we
 consider the magnetoelastic effects to be small compared to the anisotropy,  
Eq. (\ref{an}). The second term in Eq. (\ref{me}) is zero for the discussed 
states with $l_x^2=l_y^2$. The elastic energy of  tetragonal crystals 
can be written as a sum of 6 invariants:
$$F_{el}=\mu_1u^2_{xy}+\mu_2(u^2_{xz}+u^2_{yz})+
\mu_3(u_{xx}-u_{yy})^2+$$
\begin{equation}\label{E}
\mu_4u_{zz}^2+\mu_5u_{zz}(u_{xx}+u_{yy})+ \mu_6(u_{xx}+u_{yy})^2.
\end{equation}
 Minimizing  the sum of the energies, Eq. (\ref{me}) and Eq. (\ref{E}), we
find that, in the monodomain state of the unstressed crystal,
the only nonzero component of the strain
tensor is
\begin{equation}\label{u0}
u^0_{xy}=\frac{\gamma_1}{2\mu_1}l_xl_y=\pm\frac{\gamma_1}{4\mu_1}.
\end{equation}

Consider the antiferromagnetic crystal, of  plate shape, attached to
a flat elastic substrate. Let the main axis of the crystal  be
normal to the plate. The domain structure that appears in this case
is schematically  presented in Fig. 1. The state of each domain far
from the crystal-substrate boundary is one of the unstressed ground
states. Additional energetic contributions arise from i) the energy
of the domain walls, and ii) the elastic energy localized near the
crystal-substrate boundary in a layer of  thickness of the order
of the domain structure periodicity $d$ \cite{foot1}. Minimization of these two
contributions with respect to $d$ leads to the equilibrium period of
the domain structure. This argumentation  is close in spirit  
to arguments of Ref. \cite{Khach83} (see also \cite{hor91,hor04,lok}).   

In accordance with the symmetry of the domain structure, Fig. 1,
the displacement vector ${\bf u}$  has  only one component, $u_y$, 
inside the crystal, as well as in the substrate. Therefore, elasticity
equation in the crystal reduces to
\begin{equation}\label{eleq}
\mu_1\partial_x^2u_y+\mu_2\partial^2_zu_y=0.\end{equation} 
We assume that the substrate is elastically isotropic:  
 ${\mu_1=\mu_2=2\mu},$ 
where $\mu$ is the shear modulus of the substrate.
The boundary conditions at the sample-substrate contact surface are
\begin{equation}\label{bound}
u_y|_{z=+0}=u_y|_{z=-0},\\ \mu_2\partial_zu_y|_{z=+0}=2\mu\partial_zu_y|_{z=-0}.
\end{equation}

\begin{figure}
\begin{center}
\epsfig{file=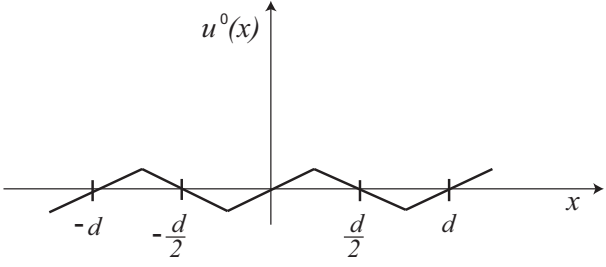, width=8cm}
\caption{The displacement field inside the crystal far from the crystal-substrate 
boundary.}
\end{center}
\end{figure}

The displacement field inside the crystal, far from the crystal-substrate boundary, 
is a zigzag function $u^0$ as schematically presented in Fig. 2. 
It's expansion in Fourier series is
\begin{equation}\label{f}
u^0(x)=\sum_{n=0}^\infty u^0_n
\sin\frac{2\pi(2n+1)x}{d}\end{equation} with coefficients 
$$u_n^0=\frac{(-1)^n\gamma_1a}{\pi^2\mu_1(2n+1)^2}.$$
It is convenient to introduce a new field $u^+$ by the relation
\begin{equation}\label{u}u_y=u^0(x)+u_y^+(x,z)\end{equation}
The Fourier series for $u^+_y$ in accordance with Eq.(\ref{eleq}) 
$$u^+_y=\sum_{n=0}^\infty u^+_n\sin\frac{2\pi(2n+1)x}{d}
\exp\left\{-\frac{2\pi(2n+1)\sqrt{\mu_1}z}{\sqrt{\mu_2}d}\right\}.$$
The Fourier series for $u_y=u^-_y$ in the substrate is
$$u^-_y=\sum_{n=0}^\infty u^-_n\sin\frac{2\pi(2n+1)x}{d}
\exp\left\{\frac{2\pi(2n+1)z}{d}\right\}.$$ 
From the boundary
conditions, Eq. (\ref{bound}), we  find
\begin{equation}\label{un}
u^+_n=-\frac{2\mu u^0_n}{2\mu+\sqrt{\mu_1\mu_2}}; \,
u^-_n=\frac{\sqrt{\mu_1\mu_2}u^0_n}{2\mu+\sqrt{\mu_1\mu_2}}.
\end{equation}

Now we can determine the total energy of the system per unit area in the $xy$-plane
\begin{equation}\label{U}
{\cal F}=-\mu_1(u^0_{xy})^2L+\nu\frac{\mu\sqrt{\mu_1\mu_2}}{2\mu+\sqrt{\mu_1\mu_2}}
(u^0_{xy})^2d+\frac{\sigma_0 L}{d},
\end{equation}
where $\sigma_0$ is the domain wall energy, and
$$\nu=\frac{8}{\pi^3}\sum_{n=0}^\infty \frac{1}{(2n+1)^3}\approx 0.27.$$
The first term in Eq. (\ref{U}) represents the energy gain due to the 
arising of the domain structure. In the monodomain state of the sample
with the free surface this is the only term of the magnetoelastic relaxation. 
The second term is the elastic energy  due
to the inhomogeneous strain near the crystal-substrate boundary. The
third  term accounts for the domain walls energy. Minimization of the 
energy, Eq. (\ref{U}),  with respect to $d$  gives the equilibrium period
\begin{equation}\label{d}
d=d_0=\left(\frac{(2\mu+\sqrt{\mu_1\mu_2})\sigma_0 L}
{\nu\mu\sqrt{\mu_1\mu_2}(u^0_{xy})^2}\right)^{1/2}.
\end{equation}

The classical law ${d\propto L^{1/2}}$, in the limit ${d\ll L}$, is the
crucial point for the experimental verification of the domain
structure theory. In the case of  antiferromagnetic domains there
exists  an additional valuable experimental possibility, namely to study the 
behavior of the structure in the external magnetic field.

The magnetic field changes the orientation of the antiferromagnetic
vector. The corresponding term  in the energy is
\begin{equation}\label{H}
F_H=-\frac{\chi_\perp-\chi_\parallel}{2}[{\bf l}{\bf H}]^2,
\end{equation}
where $\chi_\parallel$ $(\chi_\perp)$  is the
magnetic susceptibility parallel (perpendicular) to the vector {\bf
l}. In collinear antiferromagnets ${\chi_\parallel < \chi_\perp}.$

If the magnetic field ${\bf H}$ is oriented along the $y$-direction, the
minimization of the sum of the energies, Eq. (\ref{an}) and 
Eq. (\ref{H}),  gives ${l^2_x-l^2_y=h^2}$
for ${H<H_c},$ and ${l_y=0}$ for ${H>H_c}$; here ${h=H/H_c},$
${H^2_c=2\beta/(\chi_\perp-\chi_\parallel)}.$  Firstly, the magnetic field leads
to  the appearance of  homogeneous deviatoric stresses
$\sigma_{xx}=-\sigma_{yy}=\gamma_2(l^2_x-l^2_y)$ due to the second
term of the magnetoelastic energy, Eq. (\ref{me}).  These stresses
are equal in the coexisting domains, and do not affect  the
periodicity. The second effect is the renormalization of the equilibrium
strain tensor, Eq. (\ref{u0}), by the factor $(1-h^4)^{1/2}.$ The third
effect is the renormalization of the energy of the domain walls.

In order to find the domain wall structure one should take into
account the energy of the spin space inhomogeneous rotation
\begin{equation}\label{g}
E_{ex}=\frac{g}{2}(\partial_x\varphi)^2
\end{equation}
where $g$ is a constant of the exchange interaction and $\varphi$ is the rotation 
angle of the unit antiferromagnetic vector ($l_x=\cos\varphi$). The variation of the
sum of energies, Eqs. (\ref{an},\ref{H},\ref{g}), gives the
equilibrium equation
\begin{equation}\label{bg}
\delta^2\partial^2_x\varphi=(h^2+\sin^2\varphi-\cos^2\varphi)\sin\varphi\cos\varphi,
\end{equation}
where $\delta=(g/2\beta)^{1/2}$ is the effective thickness of the
domain wall. The first integral of this equation is
\begin{equation}\label{bg1}
(\delta\partial_x\varphi)^2=\left(\frac{1-h^2}{2}-\sin^2\varphi\right)^2.
\end{equation} The constant of  integration is defined by the condition far from
the domain wall where $\partial_x\varphi=0.$

\begin{figure}
\begin{center}
\epsfig{file=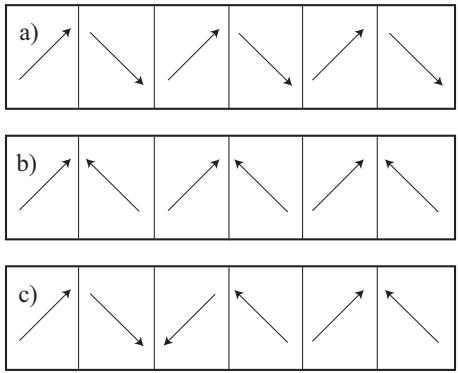, width=6cm}
\caption{Two types of ordered domain structures, a) and b) and 
 a disordered domain structure, c).}
\end{center}
\end{figure}

\begin{figure}
\begin{center}
\epsfig{file=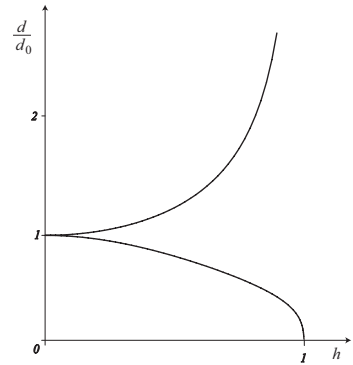, width=4.5cm,}
\caption{Dependence of the period of the domain structure on the magnetic field.}
\end{center}
\end{figure}

Finally, we find the domain wall energy
\begin{equation}\label{sigmaH}
\sigma_\pm(h)=\sigma_0\left(\sqrt{1-h^4}\pm2h^2\arcsin\sqrt{\frac{1\pm
h^2}{2}}\right),
\end{equation}
where ${\sigma_0=\beta\delta}$ is the energy of the domain wall without
 magnetic field. The minus sign  in Eq. (\ref{sigmaH})
corresponds to the domain walls presented in Fig. 3a. These walls
disappear at the critical field: 
${\rightarrow|\rightarrow|\rightarrow|\rightarrow 
\Rightarrow\,
\rightarrow\,\rightarrow\,\rightarrow\,\rightarrow\,}$
because  the difference between neighboring domains
vanishes and  the rotation in the wall tends to zero.  The
plus sign corresponds to the domain walls presented in Fig. 3b. These
walls remain at $h>1$:  $\rightarrow|\leftarrow|\rightarrow|\leftarrow.$
If the field is oriented along the $x$ direction, the behavior of the domain walls 
in Fig. 3a,b will be opposite. The energy of the walls of type $a)$ will be 
$\sigma^+$ and at the critical field we obtain the following domain structures
$\uparrow\,|\,\downarrow\,|\,\uparrow\,|\,\downarrow.$
The walls of type $b)$ will have energy $\sigma^-,$ and vanish at the critical field
${\uparrow\,|\,\uparrow\,|\,\uparrow\,|\,\uparrow\,
\Rightarrow\,\uparrow\,\uparrow\,\uparrow\,\uparrow.}$

In the general case one starts from an arbitrary distribution of the types of  walls, 
as it is presented in Fig.3c. It leads to a complicated and even
irreversible behavior of the periodicity for increasing and decreasing
magnetic fields. Nonetheless,  a simple behavior can be achieved if one
starts by applying a field higher than $H_c$, say in the 
$y$-direction. In this case, in equilibrium, one obtains a homogeneous
state (say $l_x=1$). By decreasing the field to $H<H_c$, a domain
structure will appear with energetically preferable walls, with the minus sign 
in Eq. (\ref{sigmaH}) \cite{foot2}. 
Then, the field dependence of the period is given by 
$$d^-(h)=\frac{d_0}{\sqrt{1-h^4}}\left(\sqrt{1-h^4}-
2h^2\arcsin\sqrt{\frac{1- h^2}{2}}\right)^{1/2},$$
see the lower curve in  Fig. 4. This behavior will be reversible in the
magnetic field. If then, after turning off the field, one applies a
field in  $x$-direction, the domain walls will become unfavorable,
and with increasing field the period will increase in accordance with
$$d^+(h)=\frac{d_0}{\sqrt{1-h^4}}\left(\sqrt{1-h^4}+
2h^2\arcsin\sqrt{\frac{1+h^2}{2}}\right)^{1/2},$$
see the upper curve in  Fig. 4. But, if one stops the increase of the field at
some value $H<H_c,$ and starts to decrease it, the period will decrease
more rapidly than the upper curve, because it is favorable to produce
new walls with the minimal energy. Note, that if  magnetoelastic effects 
are comparable with the anisotropy, Eq. (\ref{an}), all obtained results for 
the behavior in the magnetic field  remain valid. 
One merely should  renormalize the value of $H_c,$ and $\delta.$ 

The behavior of the domain structure will be much more complicated if
one applies a field with some arbitrary orientation. Then the
nearest-neighbor domains should have different widths. 
Let us calculate the concentrations 
of different domains, $c^+$ and $c^-$ ($c^+ +c^-=1$), in the limit of small 
magnetic fields, $H_x,H_y<< H_c$ and, as before, assuming  magnetoelastic effects to be 
small compared to the anisotropy. In this approximation we can neglect the 
small rotation  of the domain structure, which inevitably arises in the general case, 
and also set $l_x^2=l_y^2\approx 1/2$. Since the average strain is zero, we have 
$u_{yy}=0$ and $u_{xi}^+c^++u_{xi}^-c^-=0$ ($i=x,y$). At the domain wall we 
have boundary conditions of mechanical equilibrium, 
$\sigma_{xi}^+=\sigma_{xi}^-=\sigma_{xi}$ and phase equilibrium \cite{foot3}
\begin{equation}\label{phase} 
F^+-\sigma_{xx}u_{xx}^+-2\sigma_{xy}u_{xy}^+
=F^--\sigma_{xx}u_{xx}^-- 2\sigma_{xy}u_{xy}^-,
\end{equation}
where $F=F_{me}+F_{el}+F_H$; $\sigma_{xx}=\partial F/\partial u_{xx}$
and $\sigma_{xy}=\frac{1}{2}\partial F/\partial u_{xy}$ are the components of 
the stress tensor. 
Solving this system of equations, we find the concentrations  of 
the different domains:
\begin{equation}\label{concentration}
c^{\pm}=\frac{1}{2}\left(1\pm \frac{H_xH_y}{H_{me}^2}\right), 
H_{me}^2=\frac{\gamma_1 ^2}{4\mu_1(\chi_\perp-\chi_\parallel)}<<H_c^2.
\end{equation}
The domain period  cannot be found analytically at arbitrary field orientations. In
this case there are no symmetry arguments forcing  the domain walls,
near the sample-substrate boundary (at 
distances of the order of the period), 
to be flat and 
oriented along the $z$-axis. The domain walls should be inclined
and curved in this case.

Let us estimate the  parameters of  the theory. The
anisotropy energy parameter is  
${\beta\sim\alpha^4Ua^{-3}},$ where
${\alpha\sim10^{-2}}$ is the fine structure constant, ${U\sim10^4K}$
is the atomic energy, and ${a\sim10^{-8}cm}$ is the atomic size.
The constant  ${g\sim J/a,}$ where ${J\sim10^2K}$ is 
the exchange energy. Then, for the domain wall width we find
$\delta\sim(g/\beta)^{1/2}\sim a\alpha^{-2}\sqrt{J/U}\sim10^{-5}cm.$
 The shear modulus of a typical material is
${\mu\sim Ua^{-3}},$ 
 and the magnetoelastic coupling constant   
${\gamma\sim\alpha^2Ua^{-3}.}$ 
Finally, for a crystal with  thickness ${L\sim 0.5mm}$ we obtain a reasonable 
estimation of the period ${d\sim \sqrt{\delta L}}$ to be of the order of a few microns.

Note, that for  easy plane antiferromagnets with hexagonal crystal symmetry, anisotropy
effects in the plane are small $(\sim\alpha^6)$ compared to magnetoelastic 
effects Eq. (\ref{me}) $(\gamma_1=4\gamma_2).$ If we neglect this anisotropy the 
orientation of the domain structure will be arbitrary  in the absence of a  
magnetic field. 
The antiferromagnetic vector will be oriented at an angle $\pm\pi/4$ to the domain 
walls. In the presence of the magnetic field the domain 
structure will presumably  be oriented 
perpendicular to the  field \cite{foot2}.
The critical field is
$H_c\sim\gamma/\sqrt{\mu(\chi_\perp-\chi_\parallel)}$ in this case. 
The domain concentration remains 
$1/2,$ and the dependence of the period on the magnetic field 
will be described by the value of $d(h)$ obtained  
in the tetragonal case.   

This work is supported in part by the Deutsche 
Forschungsgemeinschaft (Grant No. DPP 1120), 
 by the RFBR Grants No. 06-02-16509, 06-02-17281, and by the RF President Program. 
V.I.M. thanks Forschungszentrum J\"ulich for its hospitality.  


\begin{thebibliography}{99}
\bibitem{foot1} The additional elastic energy term  
appears only  near the crystal-substrate boundary and does not appear  
near the free surface. This is a crucial difference between these elastic domains 
and the domains in ferromagnets.
\bibitem{foot2}
 The global domain
wall orientations ($x$, or $y$) remain degenerate in the
considered approximation. However, if we take into account demagnetization effects, the 
orientation of the domain structure with  walls normal to the applied field will 
be more favorable.
\end{thebibliography}
\end{document}